\RequirePackage{fix-cm}
\documentclass{svjour3}                     
\smartqed  
\usepackage{graphicx}
\usepackage{latexsym}
\begin{document}

\title{ Soliton-comb structures in ring-shaped optical microresonators: generation, reconstruction and stability 
}
\subtitle{}


\author{Rodrigues D. Dikand\'e Bitha       \and
        Alain M. Dikand\'e  
}


\institute{Rodrigues D. Dikand\'e Bitha \at
              \email{rodrigue.donald@ubuea.cm}\\ Laboratory of Research on Advanced Materials and Nonlinear Science (LaRAMaNS),
Department of Physics, Faculty of Science, University of Buea P.O. Box 63, Buea, Cameroon.        
           \and
           Alain M. Dikand\'e \at
Corresponding author, \email{dikande.alain@ubuea.cm}\\
              Laboratory of Research on Advanced Materials and Nonlinear Sciences (LaRAMaNS),
              Department of Physics, Faculty of Science, University of Buea P.O. Box 63, Buea, Cameroon.
            }

\date{Received: date / Accepted: date}

\maketitle

\abstract{
Characteristic features of soliton-comb structures in optical microresonators are investigated in normal and anomalous dispersion regimes, when the detuning parameter is varied over a broad range of values. The study rests on the assumption that soliton combs are self-organized ensemble of co-propagating coherently entangled states of light, and depending on the group-velocity dispersion they can result from space-division multiplexing of single-bright and single-dark solitons. Their analytical and numerical reconstruction schemes are discussed, while a linear-stability analysis leads to a $2\times 2$ Lam\'e eigenvalue problem whose boundstate spectrum is composed of a Goldstone-type translation mode and stable internal modes, as well as unstable decaying modes and growing modes. A power-spectral analysis of the three distinct possible soliton crystals enables us probe their inner structures in the frequency domain, and unveil the existence of structural defects in their power spectra. 
} 
\authorrunning{Dikand\'e and Dikand\'e} 
\titlerunning{Soliton crystals in ring-shaped microresonators}

\maketitle
\section{Introduction}
\label{intro}
Frequency combs~\cite{ref1} have attracted a great deal of attention in the recent past because of their numerous potential applications in optical metrology~\cite{ref2,ref3,ref41}, where they introduce the possibility to accurately measure time and frequency using optical comb structures. A frequency comb is an optical device composed of millions of equidistant modes laser that can achieve one octave-spanning spectrum. Thus an optical frequency comb can be used as an optical ruler to accurately determine any unknown optical frequency, when connected directly to a high accurate microwave or an optical clock to which the comb is stabilized to one degree of freedom. For instance the optical frequency comb can precisely measure the repetition rate frequency $f_r$ of the comb \cite{ref6}. Applications of frequency comb are increasing with progress in the field, especially in precision spectroscopy where millions of laser lines of precisely controlled combs are exploited for spectral broadband molecular spectroscopy~\cite{ref4}. Beyond enabling advances in spectroscopy, a stabilized optical frequency comb can lead to precision phase-control of the visible and near-infrared frequency spectra produced by mode-locked lasers~\cite{ref5}. This phase-control mode-locked laser is the basis of femtosecond optical frequency comb generator with regular and well-defined frequencies, and it is thus possible to determine absolutely the value of all the frequencies in a comb. Today, lasers frequency combs provide the best calibrators for astronomical sspectrograph, than conventionally used hollow-cathode lamps. It is already possible to achieve astronomical spectrographs with a short-term Doppler shift of 2.5cm/s~\cite{ref5a}. While this ability has greatly improved optical frequency metrology, atomic clocks and astronomical spectrographs, precision femtosecond optical-frequency combs also have a major impact on chip-scale technology~\cite{ref3,ref6,ref7}. Optical microresonator frequency combs generation is a very promising method for portable applications due to their advantage of low power consumption and chip-scale integration. Indeed, recent technologies using parametric frequency conversion in microcavity resonator to generate frequency comb, may provide a direct link from the radio frequency to optical domain on a chip~\cite{ref7a}.

Frequency combs can be generated via a number of mechanisms as for instance the periodic modulation of a continuous-wave (CW) laser~\cite{ref8}, the stabilization of the pulse train generated by a mode-locked laser~\cite{ref9} or by means of four-wave mixing (FWM) in optical microcavities~\cite{ref10}. For instance frequency combs can be produced from a CW laser through nonlinear optical parametric frequency conversion process in whispering gallery modes (WGM). Note that a WGM is an optical microcavity with ultrahigh quality factor $\mathcal{Q}$ that confines light, due to continuous total internal reflection in which a CW laser can coherently be converted into a set of many CW laser modes. Depending on the power and the laser detuning frequency, frequency combs in time domain can propagate as entangled states of light~\cite{ref10a}.

Cavity solitons have been reported in optical microresonators with a very high repetition frequency~\cite{ref11}. With appropriate power, a CW laser in resonance with an optical cavity can generate an ensemble of coherent states of light whose repetition rate depends on the free spectral range (FSR) of the cavity~\cite{ref12,ref13}. An interesting case arises when the process leads to a set of coherent states of microcombs, in particular when the spatial configuration of these states repeats in a uniform manner they can entangle via a space-division multiplexing (SDM), with a decay for very low values of the repetition rate into a unique entangled state of light~\cite{ref14,ref15}. For appropriate values of the repetition rate the intricate states of the space-division multiplex adopt a regular pattern, with a geometry typical of soliton crystals \cite{ref14,ref15,ref16,ref17}. The crystallization of cavity solitons in anomalous dispersion regime was investigated recently~\cite{ref16}, it resulted from the experiment that unlike previously observed ensembles of well-separated solitons, soliton crystals in optical microresonators are self-organized ensembles of independent particle-like excitations with a rich configuration space. In fact this specific type of soliton-lattice patterns introduces a new concept in the field of soliton combs, in the sense that they are tightly packed elementary solitons that interact within hole area of the resonator and exhibit novel and unique characteristics. 

The dynamics of soliton crystals is well known to be described by the Lugiato-Lefever equation (LLE)~\cite{ref21}, an equation also governing the dynamics of Kerr soliton combs in optical microresonators. Instructively the LLE is a perturbed cubic nonlinear Schr\"odinger equation (NLSE) for which the possibility of periodic patterns of multispot optical fields, formed from either time or space entanglements of equally separated identical high-intensity pulses, has been demonstrated and their one-to-one correspondence with the well-known bright elliptic soliton (ES) solution to the same equation unambiguously established~\cite{ref18}. In the context of soliton combs, ESs exhibit interesting features among which the dependence of their amplitude on both the coupled resonance width and the pulse repetition rate \cite{ref17}. Moreover the existence and stability of ESs can be determined via the pulse repetition rate of solitons: the smaller the pulse repetition rate the larger the pulse amplitude \cite{ref14,ref17}. Also the ES amplitude is inversely proportional to the cavity resonance width \cite{ref17}, meaning that the soliton envelope should broaden as the amplitude grows thereby reducing overlaps and collisions between pulses in the soliton-comb structure. \\

Although experiments have established unambiguously the possibility of several distinct soliton-lattice patterns in Kerr optical frequency-comb structures (see e.g. \cite{refa,refb,refd}), so far theoretical investigations of the mechanism of generation and stability of soliton combs in ring-shaped microresonators have focused mainly on patterns formed by spatial entanglement of bright solitons. The cases of soliton-comb structures composed of dark solitons, odd-polarity bright soliton lattice (i.e. alternating pulse and anti-pulse solitons) or odd-parity dark soliton lattices (i.e. alternating kink and anti-kink solitons), have captured no or only a very little theoretical attention. In particular the inner structures of these other possible soliton-comb structures have not been unveiled, and their stability properties are still not well understood as is the case for brigh-soliton lattice patterns  \cite{ref17}.  
 
In this work we carry out a detailed analysis of the generation, reconstruction, stability and propagation of soliton crystal-type combs in ring-shaped microresonators, in both normal and anomalous dispersion regimes where two distinct fundamental solitons are involved namely single-pulse (bright) and single-kink (dark) solitons. We lay emphasis on three distinct configurations of SDM namely the case involving identical pulses, the case of alternating pulse-anti pulse solitons and the case of alternating kink-anti kink solitons. By varying the cavity detuning and the soliton width we analyze the evolution of bright and dark soliton crystals in a ring-shaped microresonator, in the presence of two-mode amplitude noises.

In the next section (i.e. Sec.~\ref{sec2}) we introduce the LLE, and seek for its nonlinear localized and elliptic-soliton solutions in the normal as well as anomalous dispersion regimes, in the absence of perturbations. In Sec.~\ref{sec3} we consider analytical and numerical scenarios for the reconstruction of three distinct ESs in terms of soliton crystals, each with a well specified repetition rate. We estimate the minimum number of solitons in the soliton-comb train. In Sec.~\ref{sec4} and Sec.~\ref{sec5} we address the issue of spatio-temporal evolution of soliton crystals in optical microresonators, by proposing a numerical scheme based on spectral transforms to study the dynamics and power spectra of the three distinct soliton-crystal comb structures. We end the work with a summary of results and some relevant concluding remarks in Sec.~\ref{sec6}.

\section{\label{sec2}Model and Elliptic-soliton solutions}

The propagation of soliton combs in ring-shaped optical microresonators is commonly described by the master equation~\cite{ref21}:
\begin{eqnarray}
i\frac{\partial A}{\partial t}-\frac{D_2}{2}\frac{\partial^2 A}{\partial \theta^2}+g\vert A
\vert^2A=-i(\frac{\kappa}{2} +i\delta\omega)A + iF,
\label{e1}
\end{eqnarray}
where $A=A(t,\theta)$ is the envelope of the slowly-varying field, $\theta$ is the angular coordinate in the ring cavity and $t$ is time coordinate. $D_2 = c/n_0 D_1^2 \beta_2$ is the group-velocity dispersion (GVD) of the microresonator~\cite{ref12}, $\kappa$ is the cavity decay rate (linear loss or damping term), $\delta\omega=\omega_0-\omega_p$ is the pump detuning frequency, $\omega_0$ is the resonator frequency, $\omega_p$ is the laser pump frequency, $F=\sqrt{\frac{\kappa\eta P_{in}}{\hbar\omega_0}}$ is the pump intensity with $P_{in}$ the coupled power pump and $\eta$ the coupling efficiency. The parameter $g$ is the Kerr nonlinearity coefficient .

Eq.~(\ref{e1}) describes an optical cavity with Kerr nonlinearity, in which dissipative solitons propagate. For a highly pure optical cavity one can obtain a system of stable optical soliton combs. To analyze such idealized context we set $\kappa=0$ and $F=0$, such that Eq.~(\ref{e1}) reduces to the cubic NLSE:
\begin{eqnarray}
i\frac{\partial A}{\partial t}-\frac{D_2}{2}\frac{\partial^2 A}{\partial \theta^2}+g\vert A
\vert^2A&=\delta\omega A.
\label{e2}
\end{eqnarray}
Seeking for nonlinear stationary solutions to Eq. (\ref{e2}), we assume that the optical field $A(t,\theta)$ can be expressed~\cite{ref18}:
\begin{eqnarray}
A(t,\theta)=a(\theta)\,exp(i\beta t),
\label{e3}
\end{eqnarray}
where $\beta$ is the modulation frequency and $a(\theta)$ the amplitude of the field envelope assumed real. Substituting Eq.~(\ref{e3}) in Eq.~(\ref{e2}) we obtain:
\begin{eqnarray}
-\beta a -\frac{D_2}{2} \frac{\partial^2 a}{\partial \theta^2} + \gamma a^3 = \delta\omega a,
\label{e4}
\end{eqnarray}
which can be transformed to a first-integral equation i.e.:
\begin{equation}
\left(\frac{d a}{d \theta}\right)^2=-\frac{2(\beta + \delta\omega)}{D_2}a^2+\frac{g}{D_2}a^4+C.
\label{e5}
\end{equation}
The first-integral equation Eq.~({\ref{e5}}) expresses the amplitude profile $a(\theta)$ of an optical comb that has a permanent shape, determined by the value of the energy constant $C$. We first consider the physical context of localized-wave profiles in which the field envelope $a(\theta)$, either has a vanishing shape as $\theta\rightarrow \pm\infty$ such that $C=0$ or tends asymptotically to two degenerate extrema $\pm a_0$ as $\theta\rightarrow \pm\infty$. These two physical contexts give rise respectively to a bright (i.e. pulse) soliton solution: 
\begin{equation}
a_p(\theta)=a_0\,sech\left[\frac{\theta}{\ell_0^-}\right], \quad D_2<0,
\label{e6}
\end{equation}
and to a dark (i.e. kink) soliton solution:
\begin{equation}
a_k(\theta)=a_0\,tanh\left[\frac{\theta}{\ell_0^+}\right], \quad D_2>0.
\label{e7}
\end{equation}
More explicitely Eqs.~(\ref{e6}) and (\ref{e7}) describe a single-pulse soliton and a single-kink soliton respectively, both of amplitude $a_0=\sqrt{\frac{2(\beta+\delta\omega)}{g}}$ and widths $\ell_0^{\mp}=\sqrt{\frac{D_2}{\mp 2(\beta+\delta\omega)}}$ for $D_2<0$ and $D_2>0$.

When $C\neq 0$ the single-soliton solutions Eqs.~(\ref{e6}) and (\ref{e7}) become unstable, however the cubic NLSE can still admit nonlinear wave solutions. Indeed for non-zero values of $C$, Eq.~(\ref{e5}) turns to an elliptic first-order ordinary differential equation admitting three distinct elliptic-soliton solutions which are \cite{ref14,ref17,ref22}:

\begin{eqnarray}
a_{pp}(\theta)&=&\frac{a_0}{\sqrt{2-k^2}}\,dn\left[\frac{\theta}{\sqrt{2-k^2}\,\ell_0^-}, k\right],
\label{e8a} \\
a_{p\bar{p}}(\theta))&=&\frac{a_0}{\sqrt{2k^2-1}}\,cn\left[\frac{\theta}{\sqrt{2k^2-1}\,\ell_0^-}, k\right],
\label{e8b} \\
 a_{k\bar{k}}(\theta))&=&\frac{a_0}{\sqrt{1+k^2}}\,sn\left[\frac{\theta}{\sqrt{1+k^2}\,\ell_0^+}, k\right],
\label{e8c} 
\end{eqnarray}
in which $dn$, $cn$ and $sn$ are Jacobi elliptic functions of modulus $k$ ($0\leq k \leq 1$). The three Jacobi elliptic functions are periodic in their argument $\theta$ with the respective periods:
\begin{eqnarray}
\theta_p^{d}&=& 2 K\sqrt{2-k^2}\ell_0^-, 
\label{e9a} \\
\theta_p^{c}&=& 4 K\sqrt{2k^2-1}\ell_0^-, 
\label{e9b} \\
\theta_p^{s}&=& 4 K\sqrt{1+k^2}\ell_0^+, 
\label{e9c}
\end{eqnarray}
where $K=K(k)$ is the elliptic integral of first kind, and the superscripts $d$, $c$ and $s$ refer respectively to $dn$, $cn$ and $sn$.

\begin{figure} 
\includegraphics[scale=.3]{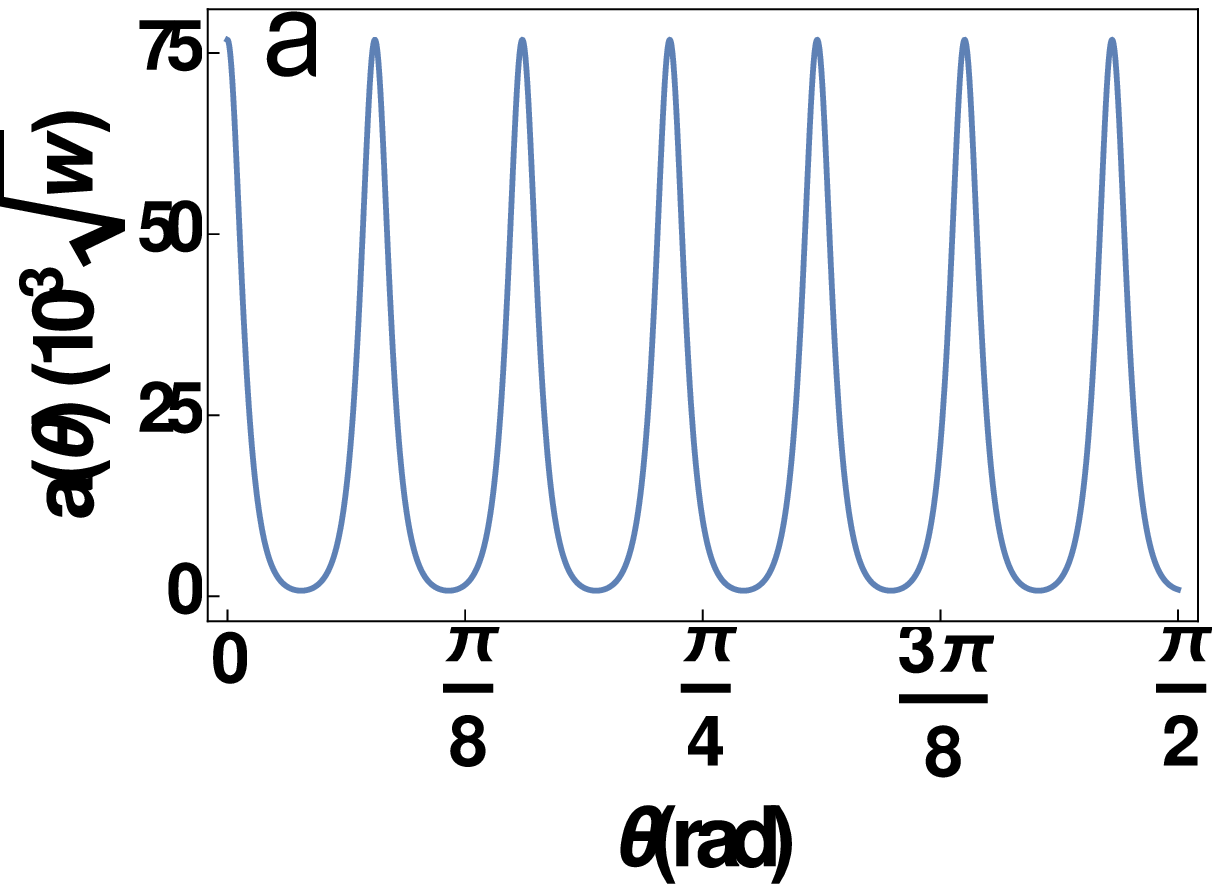}
\includegraphics[scale=.3]{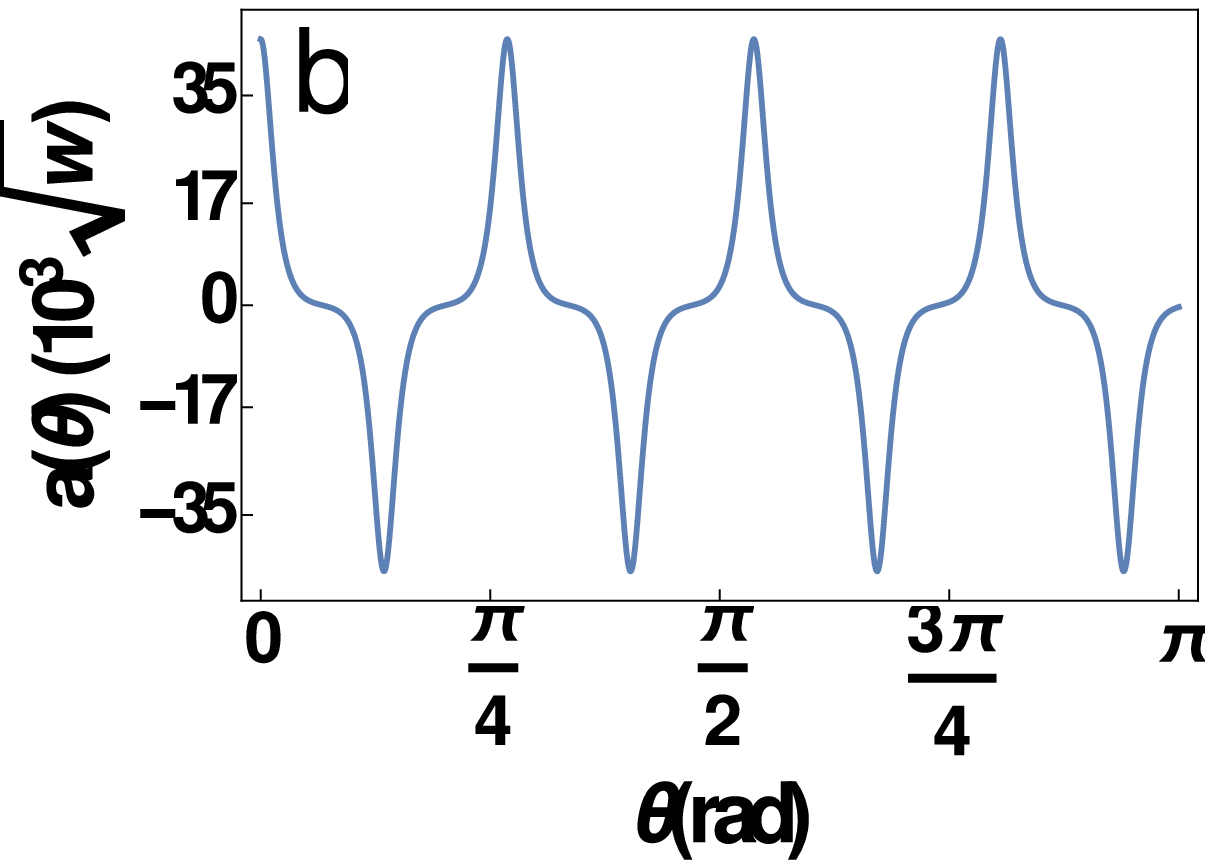}
\includegraphics[scale=.3]{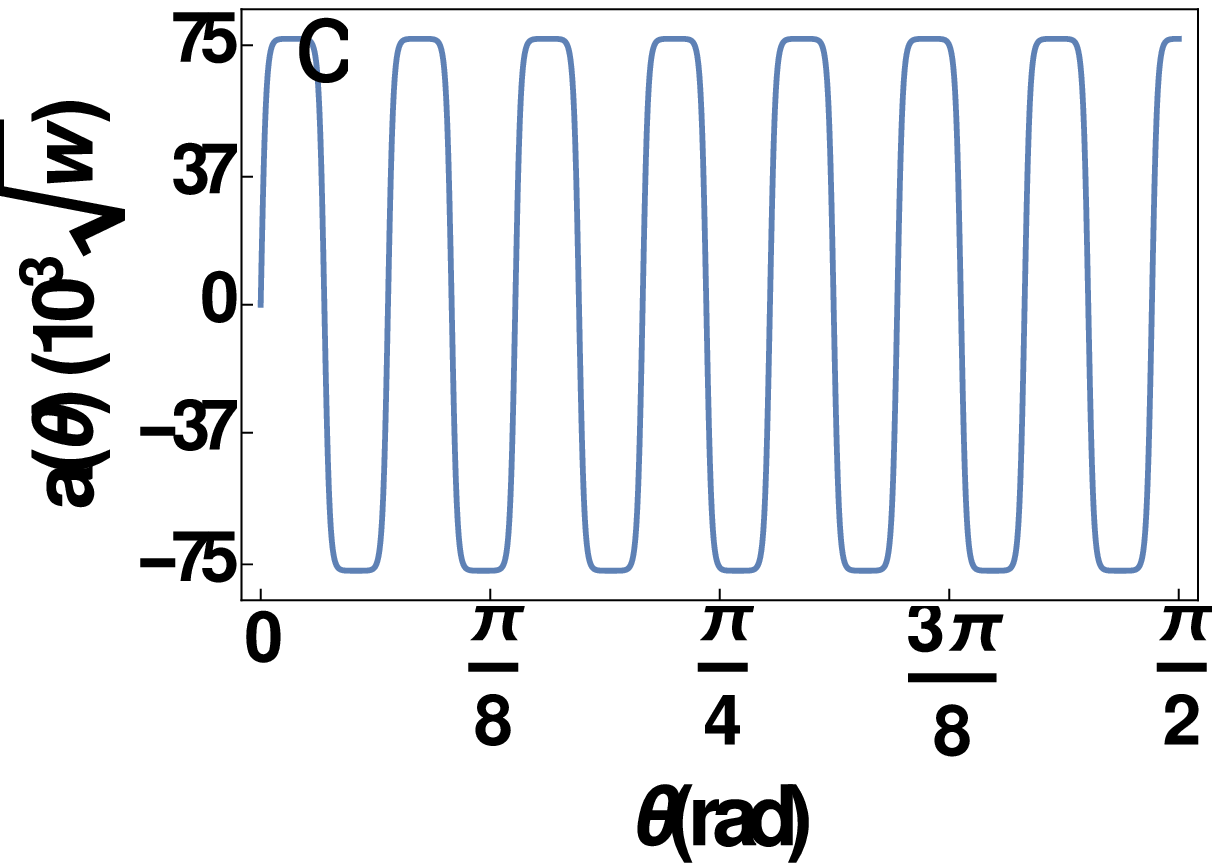}
\caption{\label{f1}
(Color online) Amplitudes of (graph \textbf{a}) the $dn$, (graph \textbf{b}) $cn$ and (graph \textbf{c}) $sn$-types elliptic-soliton solutions given by formula ~Eqs.~(\ref{e8a}), (\ref{e8b}) and (\ref{e8c}) for $k=0.97$. Graphs \textbf{a} and \textbf{b} are obtained for a negative GVD (here $D_2/2\pi=-10KHz$), and graph \textbf{c} for a positive GVD (here $D_2/2\pi=2KHz$).} 
\end{figure}

In Fig.~\ref{f1}, amplitude profiles of the three elliptic-soliton solutions Eqs.~(\ref{e8a}), (\ref{e8b}) and (\ref{e8c}) are plotted for $k=0.97$. Values of characteristic parameters are $\beta/2\pi=0.1MHz$ and $\delta\omega/2\pi=12MHz$ which correspond to typical experimental data for $MgF_2$ microresonator~\cite{ref12}.

In Fig.~\ref{f1} the $dn$ elliptic-soliton solution is reminiscent of a periodic train of evenly arranged identical pulses (graph \textbf{a}), the $cn$ elliptic-soliton solution mimics a periodic train of pulses of oppossite polarities (i.e. pulse-anti pulse) \textbf{b} while the $sn$ ellipic-soliton solution is reminiscent of a train of kink-anti kink solitons. Their shape profiles for $k=1$ coincide with those of the single-bright and single-dark soliton solutions obtained in Eqs.~(\ref{e6}) and (\ref{e7}). Instructively, when $k \rightarrow 0$ the three elliptic solitons decay into harmonic wavepackets. 

\section{\label{sec3} Reconstruction scenarios of the soliton-crystal comb structures}

It was recently established~\cite{ref10} that frequency combs could be generated in passive nonlinear microresonators. When a CW laser coupled into a WGM of high-Q pumped photons in the microcavity, broad optical spectra are generated through a cascaded four-wave mixing phenomenon. Depending on the intermodulation between the pumped photon frequencies, the FWM can lead to the generation of soliton combs in the form of formula~(\ref{e8a},\ref{e8b}) which are bright cavity solitons \cite{ref16,ref22a}, or in the form of formula~(\ref{e8c}) referred as dark cavity solitons~\cite{ref22b}.

\subsection{\label{sec3-1} Analytical Reconstruction}
Let us probe the inner structure of the three distinct ES combs obtained in formula (\ref{e8a}), (\ref{e8b}) and (\ref{e8c}). In this goal we postulate that they are in one-to-one correspondence with periodic lattices of equidistant single-soliton solutions, i.e. we can assume that the exact single-pulse and single-dark soliton solutions obtained in (\ref{e6}) and (\ref{e7}) are the fundamental components of the ES combs. This enables us envisage the following distinct SDM configurations~\cite{ref12,ref14,ref18}:
\begin{enumerate}
\item In the anomalous dispersion regime:
\begin{eqnarray}
a(\theta)&=& \sum_{J}{C_j sech\left(\sqrt{\frac{2\delta\omega}{-D_2}}(\theta-j\theta_0)\right)}.
\label{e10}
\end{eqnarray}
\item In the normal dispersion regime: 
\begin{eqnarray}
a(\theta)&=& \sum_{J}{C_j tanh\left(\sqrt{\frac{2\delta\omega}{D_2}}(\theta-j\theta_0)\right)}.
\label{e11}
\end{eqnarray}
\end{enumerate}
$j\theta_0$ and $C_j$ in the above stand for the angular position and amplitude respectively, of the $j^{th}$ fundamental soliton component in the soliton multiplex with $\theta_0$ the spatial repetition rate of single-soliton components inside the ring cavity. Note that every soliton component in the comb is an eigenfunction representing an existing mode in the soliton-comb multiplex, and obeys the same eigenvalue equation as the soliton-comb multiplex itself. We consider the specific case when all soliton components in the multiplexes Eqs.~(\ref{e10}) and (\ref{e11}) have the same normalized amplitude, given that they are all solutions to the NLSE~(\ref{e2}) we will set $C_j = a_0$ to describe the space-division multiplexing of evenly separated identical pulses, and $C_j = (-1)^j a_0$ to describe the space-division multiplexing of pulse and antipulse, or kink and antikink boundstates. In this specific case the sums in Eqs.~(\ref{e10}) and (\ref{e11}) suggest the following possible physical contexts of SDMs:
\begin{eqnarray}
   a_p(\theta)&=& a_0\sum_{J} {sech\left(\sqrt{\frac{2\delta\omega}{-D_2}}(\theta-j\theta_0)\right)},
\label{e12a} \\
a_{p\bar{p}}(\theta)&=& a_0\sum_{J}{(-1)^j sech\left(\sqrt{\frac{2\delta\omega}{-D_2}}(\theta-j\theta_0)\right)},
\label{e12b}   \\   
a_{k\bar{k}}(\theta)&=& a_0\sum_{J}{(-1)^j tanh\left(\sqrt{\frac{2\delta\omega}{D_2}}(\theta-j\theta_0)\right)}.
\label{e12c}
\end{eqnarray}
When the repetition rate $\theta_0$ is large enough~\cite{ref18} to minimize collisions between neighboring solitons, the sums in Eqs.~(\ref{e12a})-(\ref{e12c}) become exact yielding:  
\begin{eqnarray}
a_p(\theta)&=& a_0^ddn\left(\frac{\theta}{\ell_H^d},k\right),
\label{e13a} \\
a_{p\bar{p}}(\theta)&=& a_0^ccn\left(\frac{\theta}{\ell_H^c},k\right),
\label{e13b} \\
a_{k\bar{k}}(\theta)&=& a_0^ssn\left(\frac{\theta}{\ell_H^s},k\right),
\label{e13c}   
\end{eqnarray}
in which we have set:
\begin{eqnarray}
\ell_H^d&=& \frac{\pi}{2K'}\ell_0^-, \quad \ell_H^c= \frac{\pi}{4K'}\ell_0^-, \quad \ell_H^s= \frac{\pi}{4K'}\ell_0^+, 
\label{e14}\\
a_0^d&=&\frac{2K'}{\pi}a_0, \quad  a_0^c=a_0^s=\frac{4K'}{\pi}a_0,  
\label{e15}
\end{eqnarray}
where $K'=K(1-k^2)$ and $\ell_H^{d,c,s}$ is the average width of soliton components. In clear, the space-division multiplexed structures constructed in Eqs.~(\ref{e13a}), (\ref{e13b}) and (\ref{e13c}) are similar to the ES structures obtained in the previous section as solutions to the NLSE. Because of the periodic arrangement of single-soliton components in the soliton multiplexes, each of the three soliton multiplexes in Eqs.~(\ref{e13a}), (\ref{e13b}) and (\ref{e13c}) is a soliton crystal as are the ES solutions Eqs.~(\ref{e8a}), (\ref{e8b}) and (\ref{e8c}) to the NLSE. They describe the evolution of soliton combs in either anomalous or normal dispersion regimes, with the spatial repetition rates:
\begin{equation}
\theta_0^d=2K\ell_H^d, \quad \theta_0^c=4K\ell_H^c, \quad \theta_0^s=4K\ell_H^s.
\label{e16}
\end{equation}
Formula (\ref{e16}) suggests that the repetition rates $\theta_0^{d,c,s}$ are all proportional to the width of soliton ccomponents in the soliton-crystal structures Eqs.~(\ref{e13a}), (\ref{e13b}) and (\ref{e13c}), and inversely proportional to their amplitudes. Hence ES envelopes should be broadened as the amplitude grows, thereby reducing overlaps of eigenfunctions of the soliton crystals.

\subsection{\label{sec3-2}Numerical Reconstruction}

In the previous subsection we established analytically that the exact ES solutions to the NLSE were in one-to-one correspondence with artificial structures, built using the single-soliton solutions also of the same NLSE, as elementary components. We can check if the same structures are reproduced numerically, and in the course of this exercise emphasize the important role of the spatial repetition rates $\theta_0^{d,c,s}$ in fixing the appropriate number of solitons required for stable soliton crystals. To this last point, in ref.~\cite{ref23} the minimum value of the spatial repetition rate between collisionless stationary solitons, required for the formation of a stable soliton-crystal structure was defined:
\begin{equation}
\theta_{0 min}=\pi \sqrt{\frac{2\vert D_2\vert}{\delta \omega}}.
\label{e17}
\end{equation}
Knowing the minimum distance between eigenfunctions of the ESs, we can determine the maximum number of solitons $N_{max}$ that can fill the angular domain of the microresonator using the relation:
\begin{equation}
N_{max}=\frac{2\pi}{\theta_{0 min}}.
\label{e18}
\end{equation}
This quantity will be relevant in our numerical reconstruction of the soliton crystals, for this we will consider three distinct configurations i.e.:
\begin{eqnarray}
a(\theta)&=& a_0\sum_{J=0}^{N} sech\left(\sqrt{\frac{2\delta\omega}{-D_2}}(\theta-j\theta_p^d)\right),
\label{e19a} \\
 a(\theta)&=& a_0\sum_{J=0}^{N}(-1)^j sech\left(\sqrt{\frac{2\delta\omega}{-D_2}}(\theta-j\theta_p^c)\right),
\label{e19b} \\
a(\theta)&=& a_0\sum_{J=0}^{N}(-1)^j tanh\left(\sqrt{\frac{2\delta\omega}{D_2}}(\theta-j\theta_p^s)\right),
\label{e19c} 
\end{eqnarray}
where $N=2\frac{\pi}{\theta_p} \leq N_{max}$ is the number of solitons in the angular domain of the microresonator, and $\theta_p^{d,c,s}$ are spatial periods of the three ESs defined in Eqs.~(\ref{e9a}), (\ref{e9b}) and (\ref{e9c}). \\

\begin{figure}
\includegraphics[width=4.6in,height=2.3in]{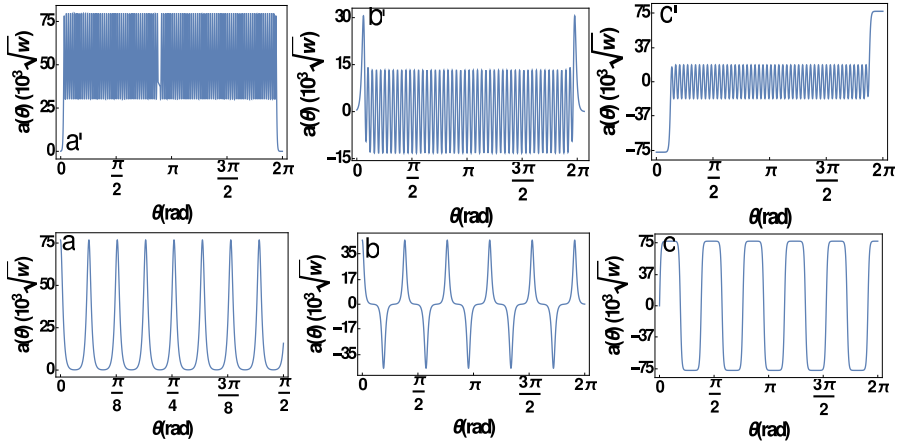}
\caption{\label{f2}
(Color online) Spatial profiles of numerically reconstructed soliton-comb structures, for small (\textbf{a'}, \textbf{b'}, \textbf{c'}) and larger (\textbf{a}, \textbf{b}, \textbf{c}) values of the repetition rates. Graph \textbf{a'}: $\theta_p^d= 0.06$ $rad$, graph \textbf{b'}: $\theta_p^c= 0.05$ $rad$, and graph \textbf{c'}: $\theta_p^s=0.05$ $rad$. Graph \textbf{a}: $\theta_p^d= 0.2$ $rad$, graph \textbf{b}: $\theta_p^c= 0.6$ $rad$, and graph \textbf{c}: $\theta_p^s=0.6$ $rad$. (note the emerging soliton-crystal patterns in \textbf{a}, \textbf{b} and \textbf{c}).} 
\end{figure}

In Fig. \ref{f2}, we plot spatial profiles of the wave structures obtained by numerical evaluation of the sums Eqs.~(\ref{e19a}), (\ref{e19b}) and (\ref{e19c}), for different values of the detuning parameter. Note that the repetition rates $\theta_p$ vary inversely with the detuning parameter, so a change in the later parameter will directly modify the number of solitons in the angular domain of the microresonator, as clearly reflected in the two figures. To better understand this behavior we stress that when $\delta \omega/2\pi=12.5MHz$, the minimum repetition rate will be $\theta_{min}=0.01\pi$ $radians$ and the maximum number of solitons in the microresonator is $N_{max}=20$. If $\delta \omega=28.2MHz$, the minimum repetition rate and the maximal number of solitons will be $\theta_{min}=0.06\pi$ $radians$ and $N_{max}=30$ respectively. However there exists a threshold value of the detuning frequency for which the number of boundstates no more allow a stable (i.e. collisionless) soliton-crystal structure. More precisely Figs.~\ref{f2} (a', b', c') show the slow-spatial-scale plots of the three SDM soliton crystals given by~(\ref{e19a}), (\ref{e19b}) and (\ref{e19c}), for $N=100$. As we can see, soliton components in the combs overlap and for each SDM we obtain a train of solitons of high-intensity amplitudes with nearly no internal structures. This last behaviour is actually due to the fact that the amplitude $a_0$ of each soliton component is proportional to the detuning frequency, while the soliton repetition rate is inversely proportional to this same parameter. Indeed, the tail of each soliton component is half the spatial period, so when the soliton tail is negligible the soliton components are no more free from collisions with neighbors. Thus we obtain a broad square-pulse signal from~(\ref{e19a}) and anti-square-pulse signal from~(\ref{e19b})~\cite{ref15} in the anomalous dispersion regime, whereas in the normal dispersion regime we obtain a broad kink signal from formula~(\ref{e19c}).

The main insight from curves of the numerical reconstruction shown Figs.~\ref{f2} (a, b, c), is that when the number of solitons in the ring cavity is reasonably large, a stable (i.e. collisionless) self-organized ensemble of co-propagating solitons forms having the crystallographic structure typical of soliton crystals. However, unlike the soliton crystals in previous works~\cite{ref16,ref17} the later ones present no defect-like vacancies and shifted pulses (or kinks), since we neglected the gain and loss in the cavity. In fact, formula~(\ref{e19a})-(\ref{e19c}) represent three sums of periodic soliton signals with periods $\theta_p^{d,c,s}$ that perfectly coincide with the numerical reconstruction of the soliton-comb structures~(\ref{e8a})-(\ref{e8c}). A close observation of  Figs.~\ref{f2}\textbf{a} and~\ref{f2}\textbf{b}, which are plots of both Eqs.~(\ref{e19a}) and (\ref{e19b}) for $D_2/2\pi=-10MHz$, reveals that they are sums of pulse signals and alternating pulse and antipulse signals respectively. Fig.~\ref{f2}\textbf{c} on the other hand describes a sum of alternating kink and antikink signals given in Eq.~(\ref{e19c}) for $D_2/2\pi=2MHz$.

\section{\label{sec4} Stability of soliton combs}
To investigate the response of ESs to small-amplitude noises, we consider a two-mode noise field propagating together with the ESs~(\ref{e8a}), (\ref{e8b}), (\ref{e8c}) i.e.: 
\begin{eqnarray}
A(t,\theta) =\{a(\theta) +[u(\theta) - v(\theta)]e^{i\omega t} + [u^*(\theta) + \nonumber \\ v^*(\theta)]e^{-i\omega t} \}e^{i \beta t},
\label{e20}
\end{eqnarray}
where $a(\theta)$ is the ES envelope, and $u(\theta)$ and $v(\theta)$ are amplitudes of the two-mode noise field having a common frequency $\omega$ ~\cite{ref17,ref25}. Inserting Eq.~(\ref{e20}) in the NLSE~(\ref{e2}) and keeping only linear terms in $u(\theta)$ and $v(\theta)$, we obtain the following set of coupled linear equations: 
\begin{eqnarray}
\left[\frac{\partial^2}{\partial \varphi^2} + 2 (a^{d,c,s})^2 + \varepsilon\right] v + \nu u &=& 0, \nonumber \\
\left[\frac{\partial^2}{\partial \varphi^2} + 6 (a^{d,c,s})^2 + \varepsilon\right] u + \nu v &=& 0,   \label{e21}
\end{eqnarray}
where $a^d=dn(\varphi)$, $a^c=k\,cn(\varphi)$ and $a^s=k\,sn(\varphi)$ are the unperturbed solitons-crystal amplitudes, 
\begin{equation}
\varphi =\rho^{d,c,s} \sqrt{(\beta+\delta\omega)/\vert D_2\vert}\theta, 
\end{equation}
is the normalized angular domain of the microresonator with $\rho^d=\sqrt{2/(2-k^2)}$, $\rho^c=\sqrt{2/(2k^2-1)}$, $\rho^s=\sqrt{2/(1+k^2)}$, $\varepsilon = [2/(\rho^{d,c,s})^2](\beta+\delta\omega)$ and $\nu = [2/(\rho^{d,c,s})^2]\omega$. \\
Eqs.~(\ref{e21}) describe a $2\times 2$ linear eigenvalue problem, with eigenmodes represented by the two-component complex vector $(u,v)$. At steady state (i.e. when $\omega=0$) the set Eqs.~(\ref{e21}) forms two independent equations given by:
\begin{eqnarray}
\left[\frac{\partial^2}{\partial \varphi^2} + 2 (a^{d,c,s})^2 + \varepsilon\right] v &=& 0, \nonumber \\
\left[\frac{\partial^2}{\partial \varphi^2} + 6 (a^{d,c,s})^2 + \varepsilon\right] u &=& 0,
\label{e22}
\end{eqnarray}
which are Lam\'e equations~\cite{ref17,ref18,ref26,ref27} of first and second orders respectively. The solutions to Eqs.~(\ref{e22}) are listed in Tables~\ref{tab:table1} and \ref{tab:table2}, together with their corresponding eigenvalues. It is interesting that the eigenmodes are all expressed in terms of ESs, in particular the eigenmode $v$ appears to be proportional to the ESs in formula~(\ref{e8a}), (\ref{e8b}) and (\ref{e8c}). The later result can be understood in that in order to remain stable, ESs must reshape the $v$ component of the amplitude noise field into another soliton crystal. This reshaping causes a unifom translation (similar to the so-called Goldstone translation) of the ESs as they interact with the noise field.  

\begin{table}
\caption{\label{tab:table1} Eigenvalues $\varepsilon(k)$ and eigenfunctions $v(\varphi)$ of the first-order Lam\'e equation~(\ref{e13a}), when $\omega=0$. $v_0^{(i)}$ are normalization constants.}

\begin{tabular}{lcr}
Eigenvalues & Eigenfunctions \\
\hline
 $(1+k^2)$ & $v^{(1)}(k)=v_{0}^{(1)}sn(\varphi)$\\
 $1$ & $v^{(2)}(k)=v_{0}^{(2)}cn(\varphi)$  \\
 $k^2$ &  $v^{(3)}(k)=v_{0}^{(3)}dn(\varphi)$\\
\end{tabular}
\end{table}

\begin{table}
\caption{\label{tab:table2} Eigenvalues $\varepsilon'(k)$ and eigenfunctions $u(\varphi)$ of the second-order Lam\'e equation~(\ref{e13b}), when $\omega=0$. $k_1^2=1-k^2$, $u_0^{(i)}$ are normalization constants.}

\begin{tabular}{lcr}
Eigenvalues & Eigenfunctions\\
\hline
 $(4+k^2)$ & $u^{(1)}(k)=u_{0}^{(1)}sn(\varphi)cn(\varphi)$\\
 $(1+4k^2)$ & $u^{(2)}(k)=u_{0}^{(2)}sn(\varphi)dn(\varphi)$  \\
 $(1+k^2)$ & $u^{(3)}(k)=u_{0}^{(3)}cn(\varphi)dn(\varphi)$ \\
 $2[(1+k^2)\mp \sqrt{1-k^2k_1^2}]$ & $u^{(4,5)}(k)=u_{0}^{(4,5)}sn^2(\varphi)-u_{0}^{4,5}$\\
    & $\times\frac{(1+k^2)\pm \sqrt{1-k^2k_1^2}}{3k^2}$ \\
\end{tabular}
\end{table}

\begin{figure}
\includegraphics[width=2.6in,height=2.3in]{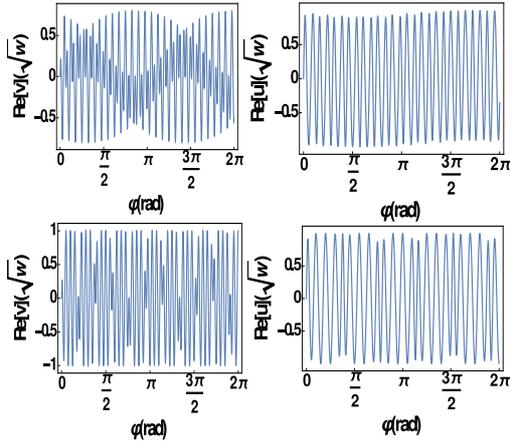}
\caption{\label{f3}
(Color online) Real parts of the internal modes v($\varphi$) (left) and u($\varphi$) (right) given in (\ref{e26}). Top: $k=0.8$, bottom: $k=97$.} 
\end{figure}

\begin{figure}
\includegraphics[width=2.6in,height=1.3in]{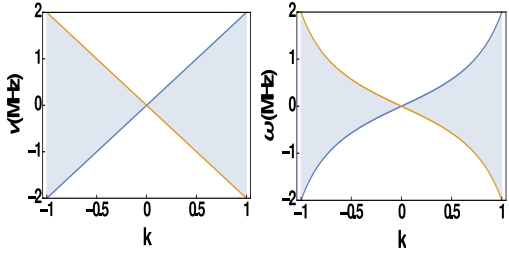}
\caption{(Color online) Plots of the internal-mode eigenvalues $\nu$ (right) and $\omega$ (left) given by (\ref{e27}),
as a function of the Jacobi elliptic modulus $k$. In the two plots the shaded region (blue region in color) denotes the stability domain.} 
\label{f4}
\end{figure}

When $\omega\neq 0$ Eqs.~(\ref{e21}) form a set of two coupled Lam\'e equations, whose solutions $\lbrack u(\varphi)$, $v(\varphi)\rbrack$ can be expanded on an orthogonal basis formed by eigenvectors $\lbrack u^{(i)}(\varphi)$, $v^{(i)}(\varphi)\rbrack$, of the uncoupled Lam\'e equations~(\ref{e22}). The solutions $\lbrack u(\varphi)$, $v(\varphi)\rbrack$ are orthonormalized according to:
\begin{eqnarray}
\int_{-K}^{K} f_j(x,k) g_p(x,k) dx = \delta_{j,p},
\label{e23}
\end{eqnarray}
where $f_j(x,k) $ and $g_p(x,k)$ are linear combinations of the eigenmodes $u^{(i)}(\varphi)$, $v^{(i)}(\varphi)$ listed in Tables~\ref{tab:table1} and \ref{tab:table2}. From the orthonormalization conditions~(\ref{e23}) we obtain the following solutions:
\begin{itemize}
 \item On the basis $\lbrack cn(x), sn(x)dn(x)\rbrack$:
 \begin{eqnarray}
v(\varphi)&=& \mp \frac{ik}{\sqrt{1-k^2}}cn(\varphi,k)e^{2i \sqrt{1-k^2} \varphi}, \nonumber 
\\
u(\varphi)&=& \left[cn(\varphi,k) + \frac{i}{\sqrt{1-k^2}} sn(\varphi,k)dn(\varphi,k)\right]
e^{2i \sqrt{1-k^2} \varphi}, \nonumber \\ \label{e24}
 \end{eqnarray}
with spectral parameters:
\begin{equation}
\varepsilon(k)=1 - 2k^2, \quad \nu(k)= \mp 2ik \sqrt{1-k^2}. 
\label{e25} 
\end{equation}
\item On the basis $\lbrack sn(x), cn(x)dn(x)\rbrack$:
\begin{eqnarray}
v(\varphi)&=& \pm sn(\varphi,k)e^{2i \varphi}, \nonumber
\\
u(\varphi)&=& \left[sn(\varphi,k) \mp i cn(\varphi,k)dn(\varphi,k)\right]
e^{2i \varphi}, \label{e26}
\end{eqnarray}
with spectral parameters:
\begin{equation}
\varepsilon(k)=1 + k^2, \quad   \nu(k)= \pm 2k.
\label{e27}
\end{equation}

Unlike in steady state where all eigenmodes are boundstates (i.e. have real spectral parameters), the last solutions show that the two-component eigenvectors $\lbrack u(\varphi)$, $v(\varphi)\rbrack$ are complex-function eigenmodes. Hence the reshaping of the two-mode amplitude noise by ESs will induce internal modes in the soliton-crystal background, costing to the ESs an energy proportional to eigenvalues of the internal modes. Internal modes with complex eigenvalues will either grow or decay exponentially and hence will be referred to as growing or decaying modes \cite{ref17}. On the other hand, internal modes with real eigenvalues can be seen as small-amplitude periodic radiations in the background of the propagating ESs. Amplitudes and eigenvalues of these internal modes are represented in Figs.~\ref{f3} and \ref{f4}.
\end{itemize}


Fig. \ref{f4} shows the variation of internal modes' eigenfrequencies with respect to the elliptic Jacobi modulus $m=k^2$. We note that as $k \rightarrow 0$ the frequencies ($\nu,\omega$) decrease to $0$. This is due to the fact that the distance~(\ref{e16}) between soliton components in the crystal decreases, thereby allowing neighboring solitons of the crystal to interact. As a consequence the soliton crystals decay into unstable wavepackets of harmonics. On the other hand, the range of allowed internal-mode frequencies increases with $k$, given that for values of $k$ close to $1$ soliton components within the train are well-spaced and therefore the soliton crystal is expected to be more stable against amplitude perturbations.

\section{\label{sec5}Numerical solutions to the LLE}

So far we have considered the generation of soliton crystals in the ring-shaped microresonator, neglecting the gain and loss considered to be perturbation terms in the LLE~(\ref{e1}). Actually, being a perturbed NLSE~\cite{ref26,ref27} the LLE is not integrable and hence only approximate solutions to the full equation can be obtained analytically~\cite{ref28,ref29}. However, numerical simulations of the equation can provide a rich insight onto the system dynamics irrespective of the orders of magnitudes of the gain and loss~\cite{ref30,ref31}. For our numerical simulations, the pseudo-spectral method (PSM) will be preferred due to its relatively better accuracy~\cite{ref32,ref33,ref34}. This method consists in mapping the evolution of the optical field in the Fourier space, where the solution of the optical system can easily be obtained with the help of a Runge-Kutta scheme. Before carrying out the simulations we need to first rewrite Eq.~(\ref{e1}) in a dimensionless form, therefore let us introduce the new variables $\tau=2t/\kappa$, $d_2=2D_2/\kappa$, $\gamma=2g/\kappa$, $\Delta\omega=2\delta\omega/\kappa$ and $\mathcal{F}=2F/\kappa$ such that the LLE (\ref{e1}) now becomes:
\begin{eqnarray}
\frac{\partial A}{\partial \tau}= -i\frac{d_2}{2}\frac{\partial^2 A}{\partial \theta^2}+i\gamma\vert A
\vert^2A -(1 +i\Delta\omega)A + \mathcal{F}.
\label{e28}
\end{eqnarray}
In order to satisfy periodic boundary conditions in accordance with the ring shape of the microresonator, we will use the fast Fourier transform (fft)~\cite{ref35}. Thus we write the dimensionless LLE~(\ref{e28}) in the Fourier space:
\begin{eqnarray}
\frac{\partial \tilde{A}}{\partial \tau}= -i\mu^2\frac{d_2}{2}\tilde{A}+i\gamma\vert \tilde{A}\vert^2\tilde{A} -(1 +i\Delta\omega)\tilde{A} + \mathcal{F},
\label{e29}
\end{eqnarray}
where $\tilde{A}(t,\mu)$ is the fast Fourier transform of $A(t,\theta)$ and $\mu$ a Fourier mode. Because the dispersion operator $\mathcal{L}=-i\mu^2\frac{d_2}{2}$ and the nonlinear operator $\mathcal{N}=i\gamma\vert \tilde{A}\vert -(1 +i\Delta\omega) + \mathcal{F}$ do not commute, the PSM will involve an error of the order $\delta t$. Therefore to shift the global error to $\delta t^2$, we will solve Eq.~(\ref{e29}) with the Runge-Kutta-Fehlberg algorithm~\cite{ref36}.

The PSM was implemented in the Matlab software and since we are interested in wavetrain-type soliton solutions describing soliton crystals, initial conditions were chosen to match the analytical expressions of the three distinct soliton-crystal structures obtained analytically in Sec. \ref{sec2}.

\begin{figure*}
\includegraphics[width=5.in,height=1.7in]{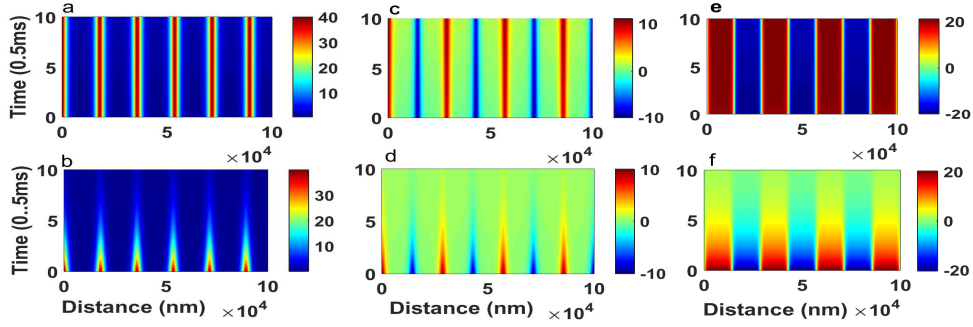}
\caption{\label{f5} Spatio-temporal evolutions of soliton crystals, obtained from numerical simulations of the LLE with $dn$ (left-column graphs), $cn$ (middle-column graphs) and $sn$ (right-column graphs) types initial conditions. \textbf{a}, \textbf{b} are for $D_2/2\pi=-10KHz$, $\delta\omega/2\pi=25MHz$ and $\ell^d=1250nm$. \textbf{c}, \textbf{d} are for $D_2/2\pi=-10KHz$, $\delta\omega/2\pi=5GHz$ and $\ell^d=1250nm$. \textbf{e}, \textbf{f}, $D_2/2\pi=-10KHz$, $\delta\omega/2\pi=22.5MHz$ and $\ell^c=1000nm$. \textbf{a, c, e} $\kappa/2 \pi= 45MHz$ and \textbf{b, d, f} $\kappa/2 \pi= 500MHz$. The unit of $A$ is $\sqrt{mW}$.}
\end{figure*}

Temporal cavity solitons are self-organized structures that arise in kerr microresonators through the double balance between dispersion and nonlinearity, and between cavity losses and external driving~\cite{ref35a}. Concerning this latter condition, the effective balance between cavity losses $\kappa$ and external driving or gain $F$ has been a challenge to ongoing experimental and theoretical works~\cite{ref35,ref35b}, giving that slight variations of the cavity losses considerably affect the lifetime of the soliton~\cite{ref16,ref17}. For illustration, in Fig. \ref{f5} we mapped the spatiotemporal evolution of soliton combs for two values of the decay rate $\kappa$. As $\kappa$ increases (from the top to the bottom rows), the lifespan of the soliton decreases exponentially with time. But since we are interested in the formation of soliton combs, we are going to consider in the rest of simulations a moderate value of the decay rate i.e. $\kappa=45MHz$.

\begin{figure*}
\includegraphics[width=5.in,height=3.7in]{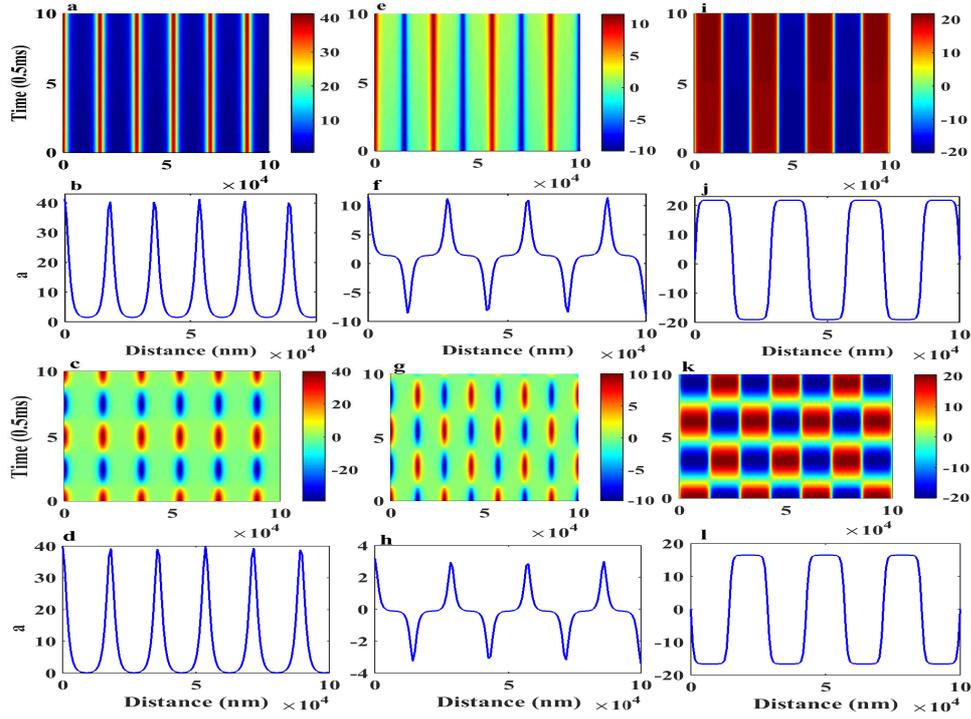}
\caption{\label{f6} Spatio-temporal evolutions of soliton crystals, with $dn$ (left-column graphs), $cn$ (middle-column graphs) and $sn$ (right-column graphs) types initial conditions. \textbf{a}, \textbf{b} are for $D_2/2\pi=-10KHz$, $\delta\omega/2\pi=25MHz$ and $\ell^d=1250nm$. \textbf{c}, \textbf{d} are for $D_2/2\pi=-10KHz$, $\delta\omega/2\pi=5GHz$ and $\ell^d=1250nm$. \textbf{e}, \textbf{f}, $D_2/2\pi=-10KHz$, $\delta\omega/2\pi=22.5MHz$ and $\ell^c=1000nm$. \textbf{g}, \textbf{h} are for $D_2/2\pi=-10KHz$, $\delta\omega/2\pi=4.5GHz$ and $\ell^c=1000nm$. \textbf{i}, \textbf{j} are for $D_2/2\pi=2KHz$, $\delta\omega/2\pi=20MHz$ and $\ell^s=1000nm$.  \textbf{k}, \textbf{l} are for $D_2/2\pi=2KHz$, $\delta\omega/2\pi=4GHz$ and $\ell^s=1000nm$. The unit of $A$ is $\sqrt{mW}$.}
\end{figure*}

In Fig.~\ref{f6}, we present the spatio-temporal evolution (top-row graphs) and spatial profiles (bottom-row graphs) of the soliton crystals in the anomalous and normal dispersion regimes, for different values of the detuning parameter. Graphs \textbf{a}, \textbf{b}, \textbf{c} and \textbf{d} in Fig~\ref{f6} are numerical solutions with a $dn$ initial profile, graphs \textbf{e}, \textbf{f}, \textbf{g} and \textbf{h} are numerical solutions with a $cn$ initial profile while graphs \textbf{i}, \textbf{j}, \textbf{k}, \textbf{l} are numerical solutions with an $sn$ initial profile. 
Besides shape profiles which are quite suggestive of the three distinct soliton crystals, we also notice a change in the soliton-crystal repetition rates as the detuning parameter is increased. The observed decrease of the repetition rates as we increase the detuning parameter is actually the result of an increase of the number of solitons in the soliton crystals, with increase in the detuning frequency consistently with previous findings~\cite{ref12,ref17}. For very high values of the detuning frequency, the repetition rate becomes small enough causing an overlap of soliton components in the soliton crystals~\cite{ref16}.

\begin{figure}
\includegraphics[width=3.in,height=2.5in]{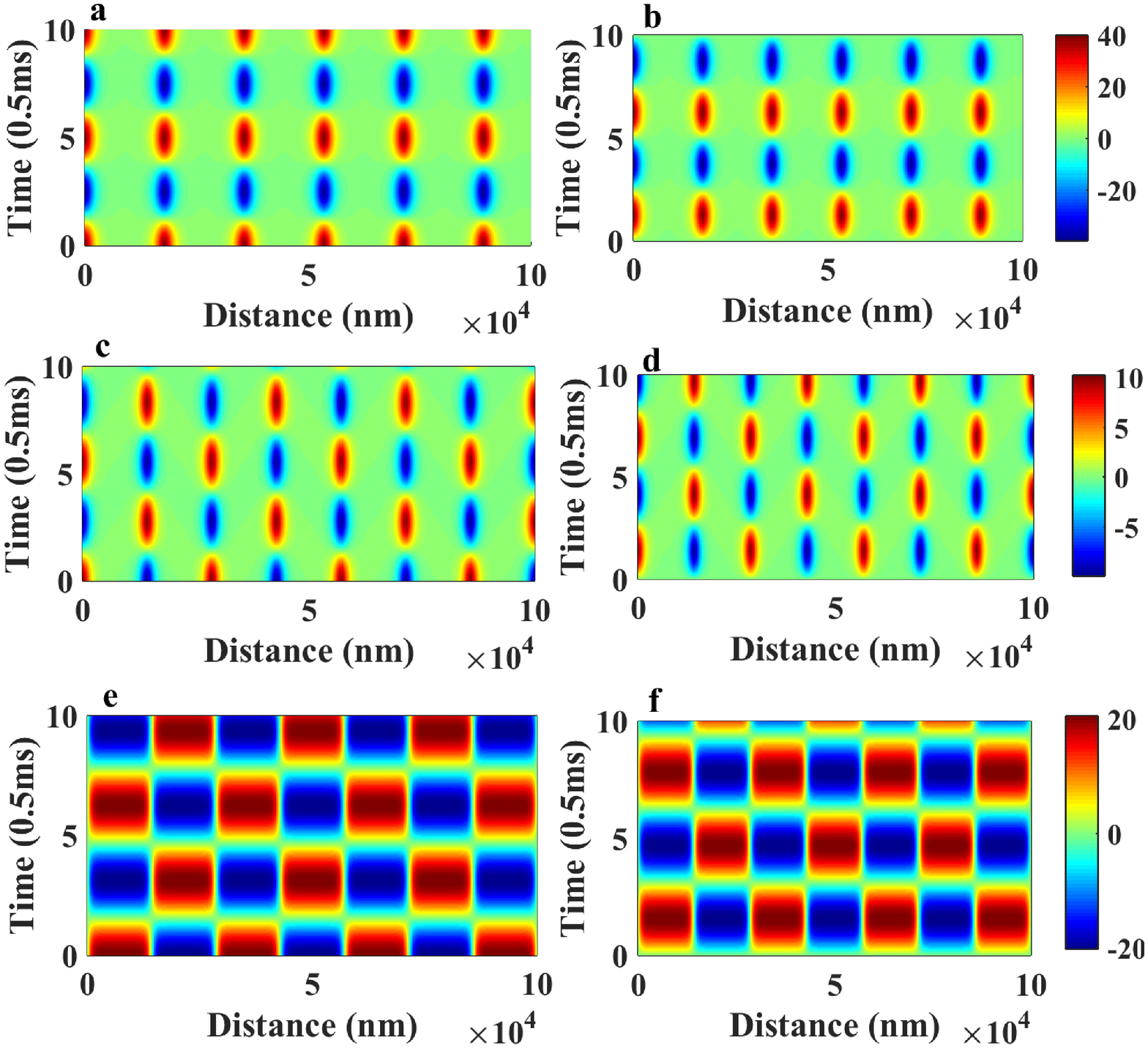}
\caption{\label{f7} Spatio-temporal evolutions of the $dn$ (top row graphs), $cn$ (middle row graphs) and $sn$ (bottom row graphs) types soliton-comb structures, when the detuning frequency is in the microwave domain: the temporal evolution of the soliton combs is similar to the evolution of an harmonic wave. The spatio-temporal evolution of the real part of the soliton-crystal comb is represented on the left, while the right side represents the evolution of the imaginary part. \textbf{a}, \textbf{b}, $D_2/2\pi=-10KHz$, $\delta\omega/2\pi=5GHz$ and $\ell^d=1250nm$. \textbf{c}, \textbf{d}, $D_2/2\pi=-10KHz$, $\delta\omega/2\pi=4.5GHz$ and $\ell^c=1000nm$. \textbf{e}, \textbf{f}, $D_2/2\pi=2KHz$, $\delta\omega/2\pi=4GHz$ and $\ell^s=1000nm$. The unit of $A$ is $\sqrt{mW}$.}
\end{figure}

An existing periodic temporal gap was noticed in the simulations, for which amplitudes of soliton components in the soliton crystals were minimal. This temporal gap can be associated with a minimal period of time of soliton oscillations in the soliton crystals, and was consistent with the minimal temporal width defined in ref. \cite{ref12} i.e.:
\begin{equation}
\Delta t_{min}=\frac{1}{\pi D_1}\sqrt{\frac{\kappa D_2 n_0^2V_{eff}}{\eta P_{in}\omega_0 c n_2}}
\label{e31}
\end{equation}
where $D_1$ is the free spectral range of the resonator. To better understand the physical meaning of this temporal gap, in Fig.~\ref{f7} the spatio-temporal evolution of soliton crystals are plotted in real (left) and imaginary (right) spaces. We see that the soliton crystals evolve in time as an harmonic wave within a period $\Delta t_{min}$, which can readily be traduced by expressing the field envelope $A(t,\theta)$ as a wave of some spatial amplitude $a(\theta)$ undergoing temporal harmonic modulations i.e.:
\begin{equation}
A(t,\theta)= a(\theta)exp\left(i\frac{2\pi}{\Delta t_{min}}t -\epsilon \kappa t \right),
\label{e32}
\end{equation}
where $a(\theta)$ represents the space-dependent amplitude of the soliton crystals and $\epsilon$ a control parameter. 

\begin{figure}
\includegraphics[width=6.in,height=2.4in]{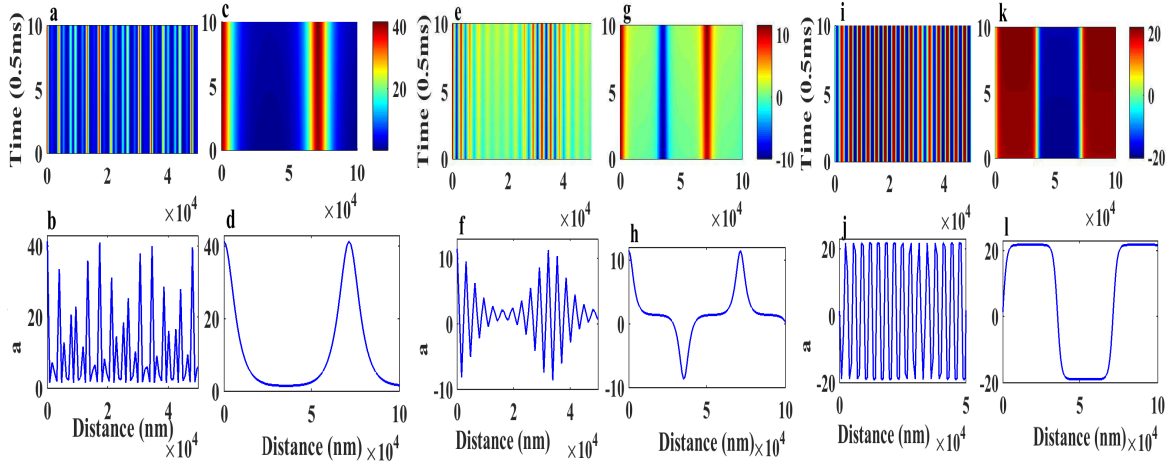}
\caption{\label{f8} Spatio-temporal evolutions (top row) and spatial profiles (bottom row) of soliton-crystal combs with the $dn$ (graphs in the two first columns on the left), $cn$ (two middle column graphs) and $sn$ (two extreme right column graphs) types initial conditons, for two different values of the soliton width: graphs \textbf{a}, \textbf{b} are for $\ell^d=125nm$. Graphs \textbf{c}, \textbf{d} are for $\ell^d=1.25pm$. Graphs \textbf{e}, \textbf{f} are for $\ell^c=100nm$. Graphs \textbf{g}, \textbf{h} are for $\ell^c=0.62pm$. Graphs \textbf{i}, \textbf{j} are for $\ell^c=100nm$. Graphs \textbf{k}, \textbf{l} are for $\ell^c=0.62pm$. }
\end{figure}    

\section{\label{sec6} Spectral analysis of elliptic-soliton combs}

To examine features of power spectra of the three soliton-comb structures, let us consider their total powers in the spectral domain i.e.~\cite{ref37}:

\begin{equation}
 Power = \vert\int A(\theta,t)e^{-i\pi\theta\mu}d\theta\vert^2,
\label{e33}
\end{equation}
where the integral is carried out over periods of the soliton combs.

\begin{figure}
\includegraphics[width=5.56in,height=3.4in]{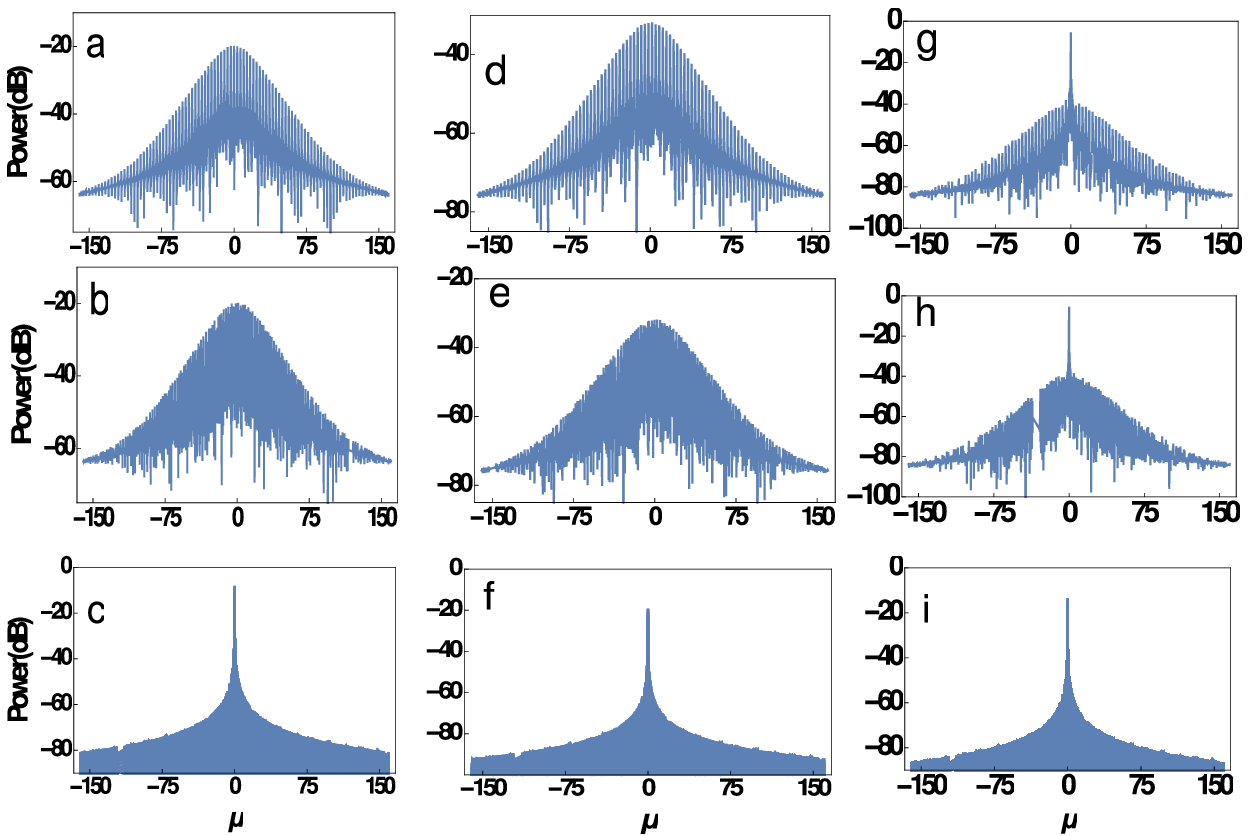}
\caption{\label{f9} Power spectra of soliton-crystal combs. Graphs from top to bottom lines were generated for three different values of the detuning frequency: line 1: $\delta\omega=5MHz$, line 2: $\delta\omega=30MHz$ and line 3: $\delta\omega=20GHz$. The power specta of columns 1 and 2 were obtained in the anomalous dispersion regime with $D_2=-10KHz$, and the plots of column 3 in the normal dispersion regime with $D_2=2KHz$. Graphs \textbf{a}, \textbf{b}, \textbf{c} are power spectra for a $dn$ soliton crystal. Graphs \textbf{d}, \textbf{e}, \textbf{f} are power spectra for a $cn$ soliton crystal. Graphs \textbf{g}, \textbf{h}, \textbf{i} are power spectra for an $sn$ soliton crystal.}
\end{figure}    

In Fig.~\ref{f9} we plotted the power spectra for the three distinct types of ESs, each graph in the figure describes a broad and discrete spectral modulations~\cite{ref38} corresponding to a frequency-comb structure representing the power spectrum of soliton crystals. From top to bottom of the figure we notice that the number of modes (i.e. discrete lines) of the spectrum increases. In fact in graphs \textbf{a}, \textbf{d} and \textbf{g} of Fig.~\ref{f9}, we can see that the internal structure of the comb is made up of evenly spaced modes. However, when the detuning frequency is increased we obtain a spectrum with nearly no internal structures (see graphs \textbf{b} and \textbf{c}, \textbf{e} and \textbf{f}, \textbf{h} and \textbf{i} for the $dn$, $cn$ and $sn$-types soliton crystals respectively). Hence the more we increase the detuning frequency or the more the number of solitons in the nonlinear wavetrains, the more dense the frequency-comb spectrum~\cite{ref12,ref13,ref38}. Nevertheless, when the number of frequency modes increases drastically the frequency comb becomes unstable with the emergence of irregular patterns reminiscent of Schottky defects (Graph \textbf{g}) or Frenkel defects (graphs \textbf{b} and \textbf{h}) observed in recent experiments~\cite{ref16}.

\section{\label{sec8} Concluding remarks}
We investigated the generation, reconstruction, stability, propagation and power spectra of three distinct soliton-comb structures in ring-shaped microresonators, considering both regimes of anomalous and normal dispersions. We established that the three distinct soliton-comb strutures were three distinct possible elliptic-soliton solutions to the NLSE, and shown that they were in one-to-one correspondences with periodic lattices of equally separated pulse, pulse-antipulse or kink-antikink soliton components. It is remarkable that the three distinct soliton-comb structures and their elementary soliton components are solutions to the same equation which is the NLSE. A linear stability analysis of the three soliton-comb structures was carried out, assuming a two-mode amplitude noise coexisting with the periodic soliton-comb structures in the microresonator. We found complex spectra of boundtstates and internal modes associated with excitations of the noise by the soliton-comb structures. A direct simulation of the LLE including the perturbation terms was performed, the aim being to check the effects of the gain and loss on the soliton crystal profiles. We found that profiles of the soliton-comb structures were preserved for reasonably small values of the loss, gain and pump field intensities. Last a power-spectral analysis was done which also unveiled a rich spectral structures for the three soliton crystals. 

Numerical results were in good agreement with previous theoretical as well as experimental works~\cite{ref12,ref13,ref14,ref15,ref16,ref17}, in particular we found that an increase of the detuning frequency induces an increase of the number of solitons in the three soliton crystals. However the soliton-comb structures are more an more unstable as the detuning is increased. We also established that the number of solitons in the soliton crystals was inversely proportional to the width of soliton components, causing defects in the soliton crystals for very small values of the soliton width. The power spectra of the three soliton crystal-type combs revealed that as the detuning frequency is increased the power spectrum becomes dense, and can display Frenkel and Schottky type defects.

\section*{Acknowledgments}
A. M. Dikand\'e thanks the Alexander von Humboldt (AvH) foundation for logistic supports. The work of A. M. Dikand\'e was done in part at the Abdus Salam International Centre for Theoretical Physics (ICTP), Trieste, Italy, within the framework of the "Senior Associateship" Scheme.\\
A. M. Dikand\'e proposed the topic, contributed to calculations and wrote the manuscript, R. D. Dikand\'e Bitha contributed to calculations and did all the simulations.  

\end{document}